\documentclass{ws-procs975x65}
\begin{document}

\title{Active Galactic Nuclei and gamma rays}

\author{Berrie Giebels}

\address{Laboratoire Leprince-Ringuet, Ecole polytechnique\\
91128 Palaiseau, France\\
E-mail: berrie@in2p3.fr}

\author{Felix Aharonian}
\address{Max-Planck-Institut f\"{u}r Kernphysik\\
69029 Heidelberg, Germany\\
E-mail: Felix.Aharonian@mpi-hd.mpg.de}

\author{H\'el\`ene Sol}
\address{LUTH, Observatoire de Paris\\
5 place Jules Janssen, 92190 Meudon, France\\
E-mail: Helene.Sol@obspm.fr}

\begin{abstract}
The supermassive black holes harboured in active galactic nuclei are at the
origin of powerful jets which can emit copious amounts of $\gamma$-rays. The
exact interplay between the infalling matter, the black hole and the
relativistic outflow is still poorly known, and this parallel session of the
$12^{\rm th}$ Marcel Grossman meeting intended to offer the most up to date
status of observational results with the latest generation of ground and
space-based instruments, as well as the theoretical developments relevant for
the field.
\end{abstract}


\bodymatter

\section{Observational context}

The detection of cosmic $\gamma$-rays is essentially based on either direct or
indirect methods. The former uses satellised high-energy particle physics type
detectors, based on the tracking of the $\gamma$-ray pair conversion and the
calorimetry of the subsequent electromagnetic cascade. This technique is
constrained almost by definition in weight, size and power consumption, which
limits respectively the maximal $\gamma$-ray energy, the effective area, and the
dynamic range. They have however a large field of view (fov), almost avoiding
any need for pointed observations since the performance of the instrument
(angular and energy resolution) changes slowly within the fov. With livetimes
close to 90\%, space-based observatories provide some of the best continuous monitoring
of the high-energy sky ever realized.

The indirect methods rely on the detection of the secondary particles induced by
the $\gamma$-ray pair creation in the atmosphere. This can be done directly on
the ground, with large area all-weather water \v{C}erenkov detectors or air
shower detectors (such as MILAGRO and ARGO-YBJ) at the price of $\sim$TeV energy
thresholds and poor background rejection. Another technique consist in the
indirect detection of the air cascade at energies $\geq10\,{\rm GeV}$, using the faint optical
\v{C}erenkov flash of the shower with atmospheric \v{C}erenkov telescopes
(ACTs), with the drawback of having a limited livetime $\sim 10\%$ due to the
same constraint as ground-based optical telescopes of clear, dark night skies
for carrying out observations limiting light curves to be continuous only over a
time up to $\sim 6{\rm h}$, with varying sensitivities and energy thresholds due
to the specifics of this technique. Since such cascades, or even the
\v{C}erenkov light pool, have transverse sizes $o(100\,{\rm m})$, the effective
area can be much larger than the physical size of the detector itself, and,
provided an efficient hadronic background, have a tremendous
sensitivity. Clearly, space- and ground-based instruments are extremely
complementary, and are very likely to remain so until much larger $\gamma$-ray
detectors can be put in space.

\subsection{Space missions}

After a long generation of $\gamma$-ray detectors based on moderately precise
spark-chambers technology running on expandable gas, such as SAS-2 (1972-1973),
COS-B (1975-1982) and EGRET (on board of \textit{CGRO}, 1991-2000), the newest
and most recent observatories \textit{AGILE} and \textit{Fermi}-LAT use robust
and finely segmented silicon-based trackers, improving the angular resolution by
an order of magnitude\footnote{For a summary of the \textit{Fermi}-LAT
  performance see {\em
    http://www-glast.slac.stanford.edu/software/IS/glast\_lat\_performance.htm}}
but also increasing the fov up to about $\pi\,{\rm sr}$ through a more compact
design. the LAT onboar \textit{Fermi} detects $\gamma$-rays in the range
$20\,{\rm MeV}$--$300\,{\rm GeV}$ where the lowest energy is limited by the
background and the highest energy mostly by statistics. The whole sky is covered
within two orbits, or $3\,{\rm h}$, with an integrated flux sensitivity above
$100\,{\rm MeV}$ of $\sim 2\times 10^{-6}\,{\rm cm^{-2}}\,{\rm s^{-1}}$ and
$\sim 4\times 10^{-7}\,{\rm cm^{-2}}\,{\rm s^{-1}}$ for respectively Galactic
and extragalactic ones, providing among the longest light curves ever
recorded\cite{tosti}\!.  An extremely important figure is that the LAT detects
the Crab nebula, a standard candle used in TeV astronomy, within a year at
$\sim100\,{\rm GeV}$\cite{fermicrab}\!, an energy usually defined as the border
between high energy (HE, $E>100\,{\rm MeV}$) and very high energy (VHE,
$E>100\,{\rm GeV}$) where ACTs become sensitive. Of crucial importance for AGN
studies evoked here is of course the LAT's uniform all-sky monitoring
capabilities along with its efficiency, allowing observers at any other
wavelength to have (almost) guaranteed GeV observations of their favourite
sources without need to apply for observations.  Within a year, the LAT has
collected $\sim 200\times 10^6$ $\gamma$-rays, 2 orders of magnitude more than
its predecessor. The first \textit{Fermi} catalog\cite{fermicat} (1FGL), based
on 11 months of scientific operations, contains 1451 sources of which 1043 are
at Galactic latitudes $b>10^\circ$. From this latter sample was then derived the
first LAT AGN catalog (1LAC), comprising 300 BL Lac objects (compared to 13 for
EGRET), 296 flat-spectrum radio quasars (FSRQs, of which EGRET saw 66), and 113
AGN of other (or unknown) type. This is a tremendous change of the HE sky with
just one of the 10 expected years of scientific operations, so even more is to
be expected in the future.

\def\figsubcap#1{\par\noindent\centering\footnotesize(#1)}
\begin{figure}[t]%
\begin{center}
  \parbox{2.1in}{\epsfig{figure=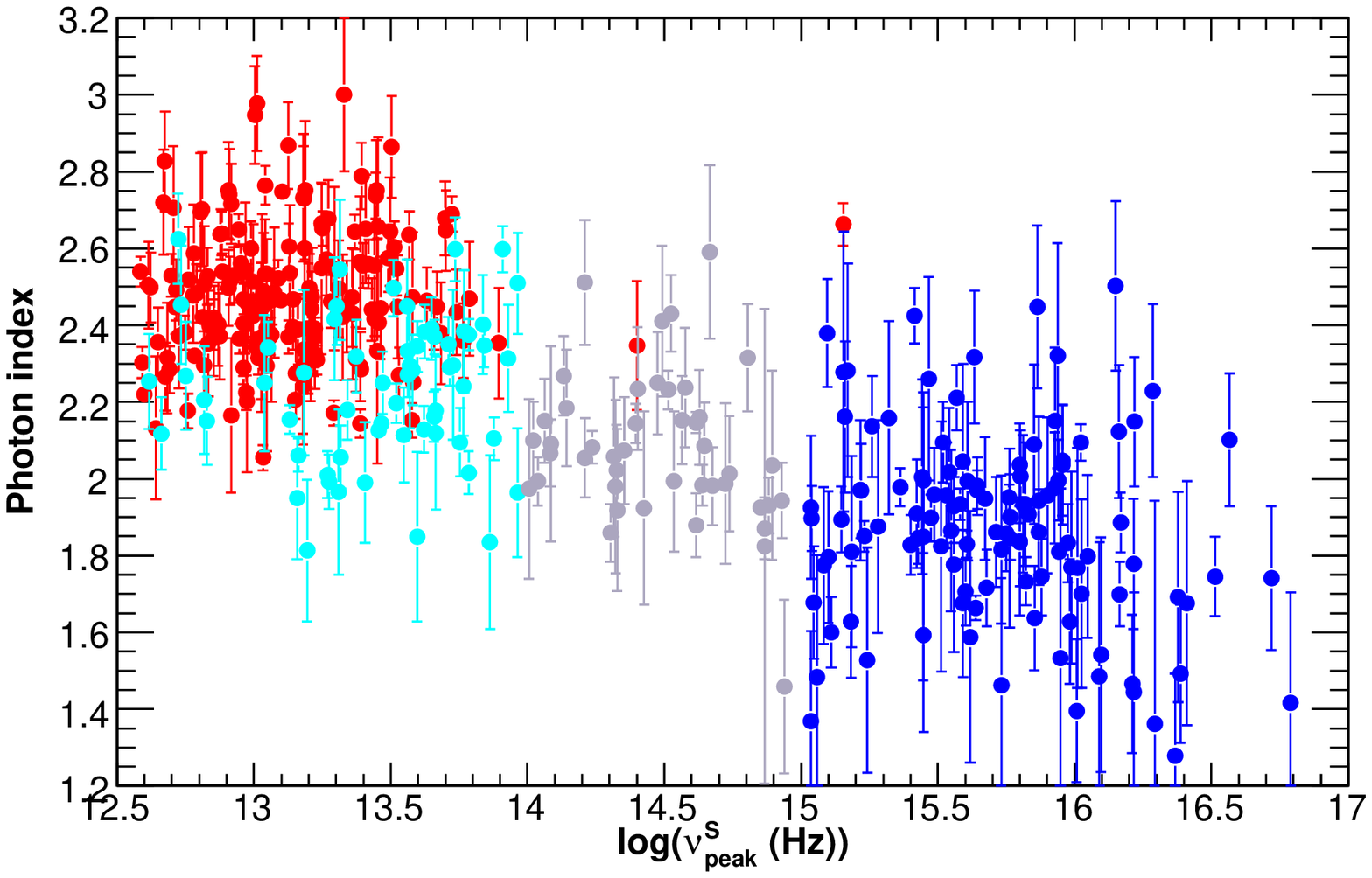,width=2.5in}\figsubcap{a}}
  \hspace*{1pt}
  \parbox{2.1in}{\epsfig{figure=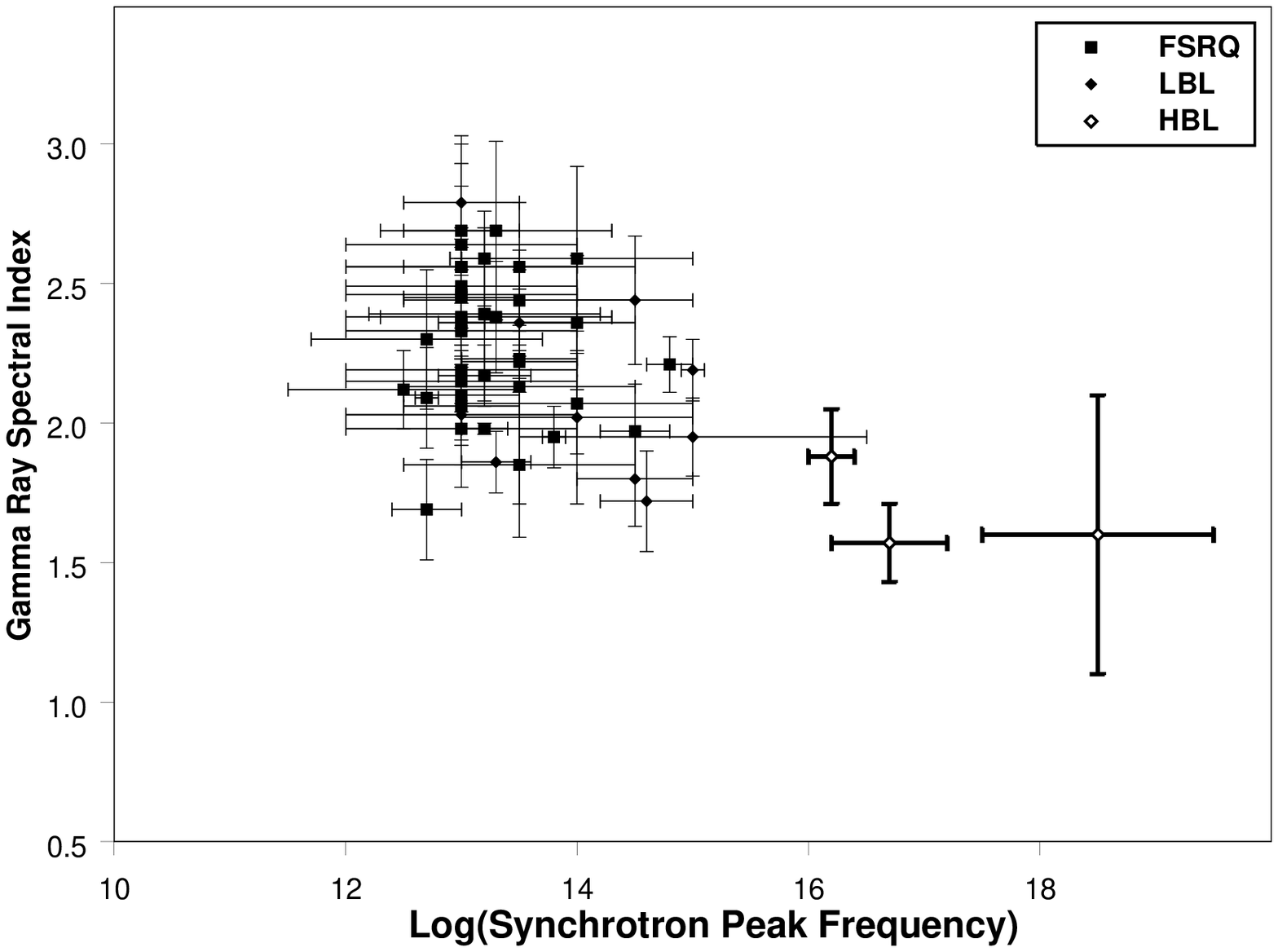,width=2.5in}\figsubcap{b}}
  \caption{(a) The \textit{Fermi} blazar photon index as a function of its derived
synchrotron peak frequency $\nu_{\rm peak}^{\rm s}$, as derived in
Ref.~\refcite{fermi1lac}. (b) A similar study using the EGRET data, as published
  in Ref.~\refcite{nandik}\!. Thanks to a much larger population and a better
  energy resolution, a clear correlation appears where the blazar spectrum
  becomes increasingly harder when the synchrotron peak $\nu_{\rm peak}^{\rm s}$
  moves blueward from the optical to the X-rays.}
  \label{fig1}
\end{center}
\end{figure}

\subsection{Ground-based telescopes}

The most sensitive extensive air shower (EAS) detector to date is the Milagro
observatory, which has provided a 1 to 100 TeV survey of the $\gamma$-ray sky in
which 34 \textit{galactic} \textit{Fermi} sources were
found\cite{milagro}\!. So far however ACTs have been the dominant technique for
studying the VHE extragalactic field, since the $\gamma$-ray horizon at TeV energies
is (at best) located at $z\sim0.1$ due to photon-photon interactions with the
extragalactic background light (EBL).

\begin{figure}[t]
\begin{center}
\psfig{file=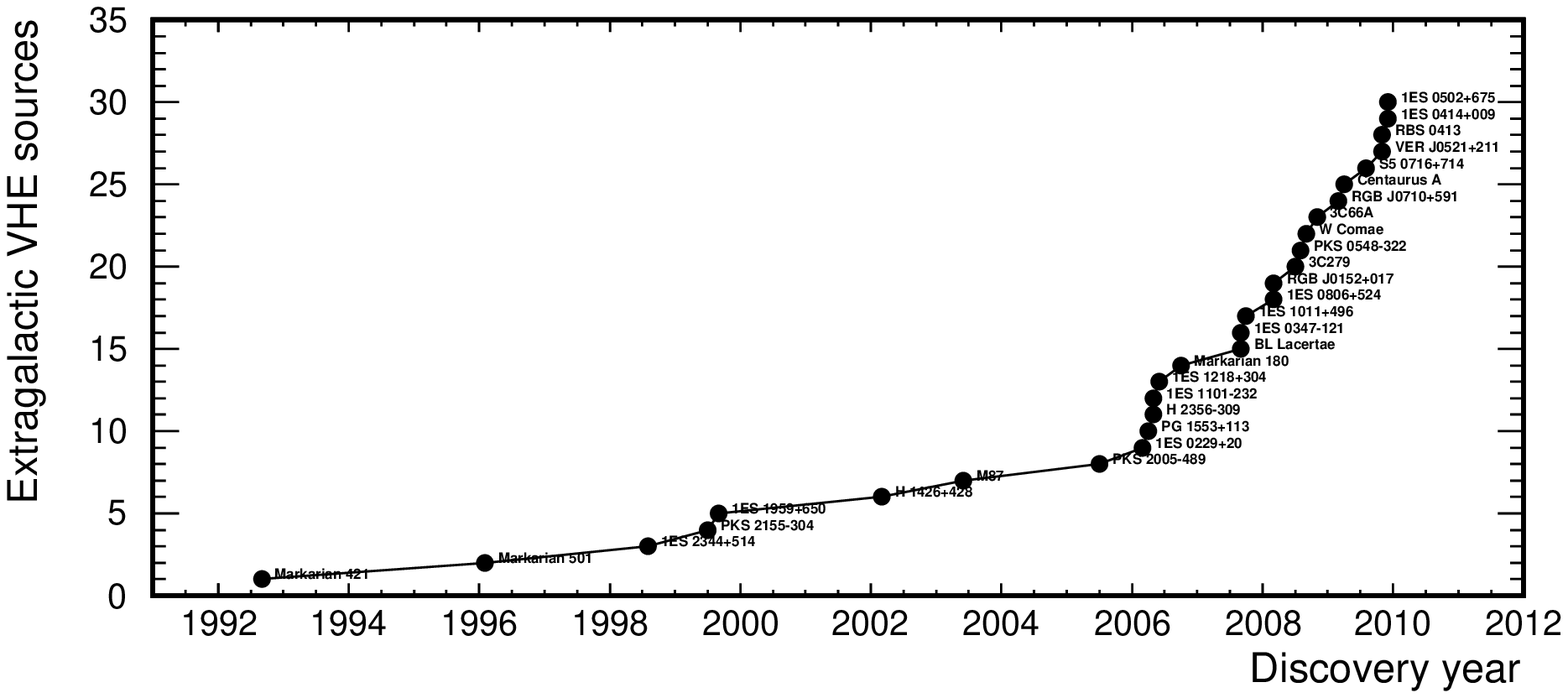,width=5in}
\end{center}
\caption{The number of extragalactic sources detected by ACTs, as a funcion of
  their detection date, using the TeVCaT. The slope change from $\sim 0.5/{\rm
    year}$ to $\sim 5/{\rm year}$ coincides with the current generation of ACTs
  becoming operational, illustrating the dramatic effect of the sensitivity
  increase.}
\label{fig2}
\end{figure}

The new generation of major ACTs (HESS, MAGIC and VERITAS mainly), with larger
telescopes and improved reconstruction techniques, have augmented the flux
sensitivity of this technique by an order of magnitude within a decade, reaching
a significance of 5 standard deviations on Crab nebula type sources in $\sim
30\,{\rm s}$ or less, and ways are found to decrease the effect of moonlight in
order to increase the observation time. The population of extragalactic ``TeV
sources'' is currently at 30, but continuously increasing\footnote{See the
  catalog TeVCat http://tevcat.in2p3.fr}\!,as can be seen in
Fig.~\ref{fig2}. Most of the sources are BL Lacs with fluxes at the level of a
few percent of the Crab nebula equivalent, with little measurable variability.

\section{Gamma-ray AGN}

While it is widely accepted that blazar $\gamma$-ray emission originates from
inverse Compton emission in an ultrarelativistic jet pointing within a few
degrees of our line of sight, it is still not clear where in the jet this
emission occurs, nor what the source of the scattered soft photons is. Beyond
these questions, it has also yet to be established what role the central black
hole exactly plays in the mechanism that powers the jet. The extremely efficient
release of binding energy of accreting gas, as well as directly extracting spin
energy from the black hole through the Blandford-Znajek mechanism\cite{bz}\!,
are possible. The newly available means to perform $\gamma$-ray multi-wavelength
monitoring of the nonthermal emission will certainly provide clues for a better
understanding about the jet composition, and perhaps about jet formation as
well. The spectral energy distribution (SED) of blazars in $\nu F_\nu$
representation usually has two broad components, with the lower one generally
believed to be synchrotron radiation, and the higher one inverse Compton (IC)
radiation from the same electrons on target photons which can have different origins.

At the time this review is being written, all but 4 of the grand ensemble of
extragalactic $\gamma$-ray sky sources are seen by \textit{Fermi}, so the global
HE spectral characteristics can be well summarized by the 1LAC
sample. Fig.~\ref{fig1}(a) shows that the HE $\gamma$-ray index $\Gamma_{\rm
  HE}$ becomes harder when the synchrotron peak shifts globally blueward. This
should be compared with Fig.~2 in Ref.~\refcite{nandik}, displayed in
Fig.~\ref{fig1}(b), where this trend was first seen using only EGRET blazars,
but with 37 FSRQs and 13 BL Lacs. It appears hence that the FSRQ $\gamma$-ray
spectra are different from the BL Lac population, with the average photon index
of the former being $\geq 2.2$, while the average photon index of the latter is
$<2$. The blazar index versus luminosity characteristics as seen by
\textit{Fermi} appear also to largely follow the so-called blazar
sequence\cite{ghi98,fos98}, based on a one zone, homogeneous synchrotron
self-Compton and external Compton model, where the peak luminosities of the two
radiative components are located at increasingly higher frequencies when the
observed $\gamma$-ray luminosity decreases, and where the external Compton
cooling becomes progressively more inefficient than the self-Compton.

On the other side, the \textit{observed} VHE spectra from all extragalactic objects have photon
indexes $>2$, so the combination of \textit{Fermi} and ACTs constrains for
almost all BL Lacs the peak of the observed SED (in $\nu F_\nu$
representation)\cite{fermigevtev}\!. This should now allow the best
characterization ever made of both components of the SED, the monitoring with
unprecedented accuracy of how these components evolve, and how they are related.

\subsection{FSRQ}

The FSRQ class of blazars are the most powerful $\gamma$-ray emitters as well as
the most luminous. This is best illustrated by the gigantic flare\cite{atel}
exhibited by 3C 454.3 on December 2, 2009, when it reached a flux of
$F(E>100\,{\rm MeV})\simeq 2\times 10^{-5}\,{\rm cm^{-2}}\,{s^{-1}}$, which is
up to date the brightest transient ever detected in the HE range. For
comparison, the brightest persistent source in the HE sky is the Vela pulsar, at
about half that flux. This population is also extremely variable, and their
spectra show significant spectral curvature in the HE range. Since
\textit{Fermi} probes these sources mostly at their peak luminosity frequency,
and with unprecedented resolution, theorists working on FSRQs are guaranteed to
have a field day. The detection of the FSRQ 3C279\cite{3c279} at $z=0.536$
during an optical flare in 2006 by the MAGIC experiment at the $5.8\sigma$ level
and a dedicated low-energy analysis was the first of this kind with an ACT. The
corresponding GeV flux, had \textit{Fermi} been there to see it, would have been
a really exceptional state, an order of magnitude higher than the highest GeV
flux seen so far on this source, as is pointed out in Ref.~\refcite{fermigevtev}
in order to match the \textit{Fermi} extrapolation to higher energies assuming a
conservative EBL attenuation and no further spectral steepening. This detection
is clearly paving the way for upcoming ACTs with $<100\,{\rm GeV}$ thresholds,
which could probe the FSRQ Compton component variability at timescales not
accessible to \textit{Fermi}, and eventually find out whether they can exhibit
very fast variability as well.

\subsection{BL Lacertae}

The most constraining observations on jet kinematics and the $\gamma$-ray
emitting region probably come from AGN observations with ACTs, mostly the
high-frequency peaked class of BL Lac objects (HBLs), which can exhibit
minute-timescale variability in their VHE $\gamma$-ray fluxes, as has been shown
by the HESS and MAGIC experiments in PKS~2155-304\cite{aha07} and
Mkn~501\cite{alb07}\!, respectively.  These findings have given raise to
considerable theoretical developments in the literature\cite{sol,levinson}\!, as
existing models have to struggle to either remain simple but invoke
uncomfortably large bulk Lorentz factors of 100 or higher in order to allow
$\sim\,{\rm TeV}$ $\gamma$-rays to escape from the compact area, or use multiple
emitting zones, or the emission region becomes very
small\cite{levinson}\!. Intruigingly, the power density spectrum of one of the
giant flares of PKS 2155-304 shows no sign of high-frequency cutoff in the red
noise power law distribution up to the Poisson noise level, so it
cannot be excluded that even faster variability could be detected in the light
curves if they were sampled by more sensitive instruments (or if the fluxes were
higher).

\begin{figure}[t]
\begin{center}
\psfig{file=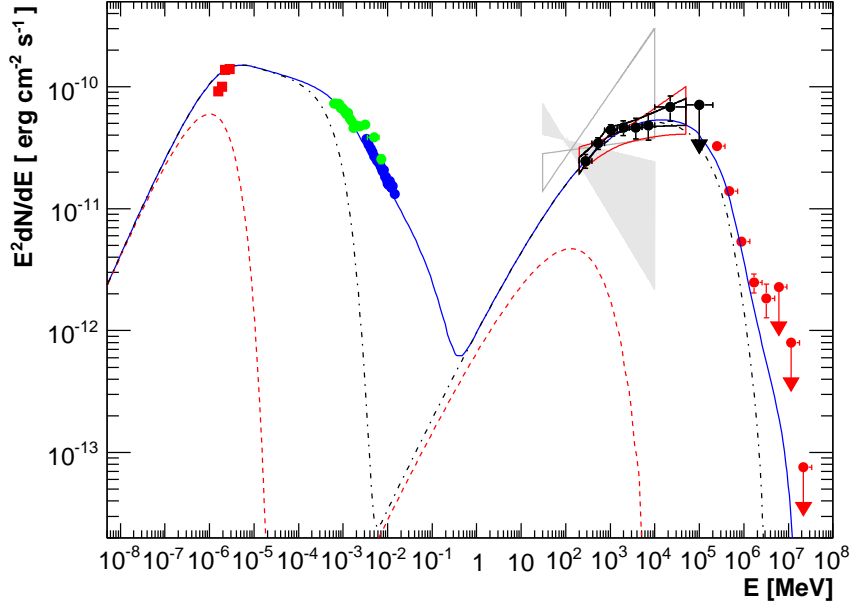,width=5in}
\end{center}
\caption{The spectral energy distribution of PKS~2155-304 during the 2009 MWL
  campaign, involving optical measurements from ATOM, X-ray observations carried
out by both \textit{Swift} and \textit{RXTE}, HE and VHE $\gamma$-ray
observations from \textit{Fermi} and HESS, respectively. The archival EGRET
measurements appear in grey. The plain, dotted, and dash-dotted lines are issued
from the same synchrotron self-Compton (SSC) model but with different electron cutoff energies, as
described in Ref.~\refcite{fermimwl}, and they show that low (or quiescent) flux states can
accomodate the absence of correlated X-ray/VHE variability (see text).}
\label{fig4}
\end{figure}

The BL Lac object PKS~2155-304 was also the first source of this type to be
targeted during a simultaneous space-ground multi-wavelength campaign in 2009
involving \textit{Fermi}\cite{fermimwl}\!. Observing the soft spectrum of the
archival EGRET measurements on this object would have been quite disconcerting
since the this soft spectrum is incompatible with a smooth connection to the
lowest known VHE spectrum (Fig.~\ref{fig4}) and would therefore rule out a
single radiative IC population. The \textit{Fermi} observations during this
campaign turned out to be actually in good agreement with the simultaneous VHE
spectrum, and provided the first simultaneous IC characterization of an HBL. An
interesting feature of this campaign was that the X-ray and VHE light curves,
despite significant variability, were not correlated, when the VHE and optical
fluxes appeared to be correlated to some degree. While the latter correlations
have yet to be confirmed and understood, the observed X-ray/VHE behaviour could actually
be reproduced with a simple SSC model since the highest electrons, those
radiating the X-rays through the synchrotron mechanism, are barely noticed at
VHE energies due to a combination of large Klein-Nishina suppression, a low
density of target photons, and EBL attenuation.

With the advent of quasi-continuous, well-sampled observations of flaring BL~Lac
objects, VHE $\gamma$-ray light curves can now be characterized beyond finding
the fastest doubling/halving time scales. Tools commonly used in X-ray
observations can now usefully be applied to the \textit{whole} data set instead
of picking the most interesting part of it (or at least what might look as
such). In this way it has now been established that, at least for PKS~2155-304,
the 2006 flaring state is compatible with a stationary lognormal
process\cite{deg08} whose power spectum density is red noise, sharing a
characteristic of accreting sources shining with thermal radiation such as Cyg
X-1 or some Seyfert galaxies for instance. This might be important enough that
it should be sought for not only in the other jet-dominated non-thermal
emissions from AGN, but also again in the same source to find out whether the
PSD slope or lognormal factor are subject to changes. Its eventual ubiquity
would also have to be reproduced by time-dependent emission models.

Not much was known about BL Lac objects in the HE range, so it was not clear
until \textit{Fermi} started observing what exactly was happening in that
regime, i.e. whether BL Lacs were just very faint, or if they were bright at
higher energies where EGRET's sensitivity dropped. It was hence surprising to
find many hard BL Lac objects among the bright sources seen by
\textit{Fermi}\cite{fermigevtev}\!. Their spectra are generally power laws over
3 decades in energy, with sometimes small deviations at energies $\geq1\,{\rm
  GeV}$. The overall SED is rather unsurprising, as the high end of the HE and
the low end of the VHE spectra tend to be well within range, and in some cases
even slightly overlap, since a few of the hardest sources with HE indexes of
$\sim1.5$ have a spectral bin as high as $100\,{\rm GeV}$. A rather special case
is the HBL PG 115+113, of unknown redshift, which has the largest spectral
difference ($\Delta \Gamma \sim 3$) between the HE and the VHE regimes. Since
its distance is still unknown, it cannot be ruled out that the observed large
spectral break is not related to the source but to the $\gamma$-ray
propagation\cite{fermi1553}\!. This source, like many other BL Lacs visible in
both the HE and VHE ranges, exhibits surprisingly little variability in either
energy band\cite{fermi1553} over long time scales when it is actually known to
be largely variable at X-ray and optical wavelengths.

There are high hopes that \textit{Fermi}, given its intra-day overall sky
surveillance and the fact that it shares the same radiative population as VHE
instruments, will prove to be a more efficient provider of flaring states in
blazars of interest for ACTs than all-sky X-ray monitors such as the ASM onboard
\textit{Rossi-XTE} or the BAT onboard \textit{Swift}. But looking at the rich
Astronomer's Telegram history of \textit{Fermi} the very variable extragalactic
sky is so far essentially composed of FSRQs and a low-frequency peaked BL Lac.

\subsection{non-blazars}

Unlike most blazars, high frequency VLBI can probe the jets in a few selected
VHE sources like M87 and Centaurus A\cite{ahacena} down to size scales ~100
gravitational radii\cite{markos}. Getting closer to
localise the source of the $\gamma$-ray emission. These sources are widely
supposed to be the unbeamed parent population of blazars, so even observations
at large angles of the jet can provide valuable insights of jet physics. in M87,
it was possible to establish that the acceleration and the collimation of the
jet occurs within $\sim 100\,R_{\rm S}$ (where $R_{\rm S}=2GM/c^2$ is the
Schwarzschild radius of the central black hole, a very important size in such
systems). A joint ACT campaign, along with VLBA observations, has revealed that
an increase of the nucleus radio flux might be the lagged counterpart of a
similar VHE $\gamma$-ray transient, implying that the $\gamma$-ray emission, and
hence the acceleration of the underlying radiative population, is likely to
happen well within the collimation region\cite{science}\!. Scenarios to explain
how $\gamma$-ray variability can be detected in off-axis jet systems include
e.g. a multiblob SSC model similar to those used for blazar
emission\cite{lenain}, a two-zone spine-sheath layer mechanism\cite{ghim87}, and
pulsar-type acceleration due to centrifugally accelerated electrons in a
rotating jet magnetosphere\cite{rieger}. Unifications of these scenarios with
the beamed population is going to be an interesting development to follow in the
future, as well as its long-term HE characterization by \textit{Fermi}. 

Another possible emerging population of $\gamma$-ray emitting AGN unseen before
\textit{Fermi} are the narrow-line Seyfert-1 galaxies (NLS1) of which up to now
4 have been detected\cite{seyfert}\!. It was not completely unexpected that the
rather rare radio-loud sample of NLS1 galaxies could be also high-energy
emitters since in some scenarios they are probably have a pole-on
orientation\cite{rem}\!. The HE spectra are however extremely steep so, given
their distance, nothing is expected to be visible by ACTs by extrapolating the
\textit{Fermi} spectra, but suprises can arise given the rather complex SEDs
which leave probably more room open to radically different models than there is for BL
Lac models.

\section{Propagation effects}

\subsection{Extragalactic Background Light}

Photons of energy $E_\gamma$ can create an electron/positron pair when colliding
with target photons of energy $E_t$ provided that $E_\gamma E_t \geq 2m^2_ec^4$,
which makes photons in the $100\,{\rm MeV}$--$100\,{\rm TeV}$ range sensitive to
various existing cosmic backgrounds\cite{ress} and especially those in the far
UV - far IR range (usually called the \textit{extragalactic background light},
or EBL), since they have the relevant density to adversely affect the VHE
extragalactic spectra when the sources are located at cosmological
distances. Due to difficulties in subtracting local galactic foregrounds, this
part of the cosmic background is poorly enough constrained that $\gamma$-ray
attenuation can be a useful tool to probe the EBL at energies despite all the
uncertainties in the intrinsic emitted spectrum\cite{naturr}. BL Lac objects can
be used for this since the radiative particle distribution is expected to be
close to the injection function, for which theoretical constraints exist,
because the Lorentz factor $\gamma_{\rm cool}$ at which electrons cool within
the dissipative region is expected to be large enough\cite{ghirecent}\! that the
radiative electron spectrum is not affected within a time $\sim R/c$. The use of
such constraints turn out to be so far the most efficient way to constrain the
EBL and hence properties of star formation rate and early Population III
stars\cite{coppi,naturr,raue}\!. The FSRQ class, despite the advantage of being
at larger distances on average than BL Lacs, might be more tricky to use for
such purposes, since not only much faster cooling time scales deviate the
radiative spectrum faster from the injected one, but also because significant
$\gamma$-ray absorption is expected inside the source
itself\cite{costa}\!. Another original use of EBL attenuation is the possibility
to constrain the Hubble constant\cite{gorecki}\!.

\begin{figure}[t]
\begin{center}
\psfig{file=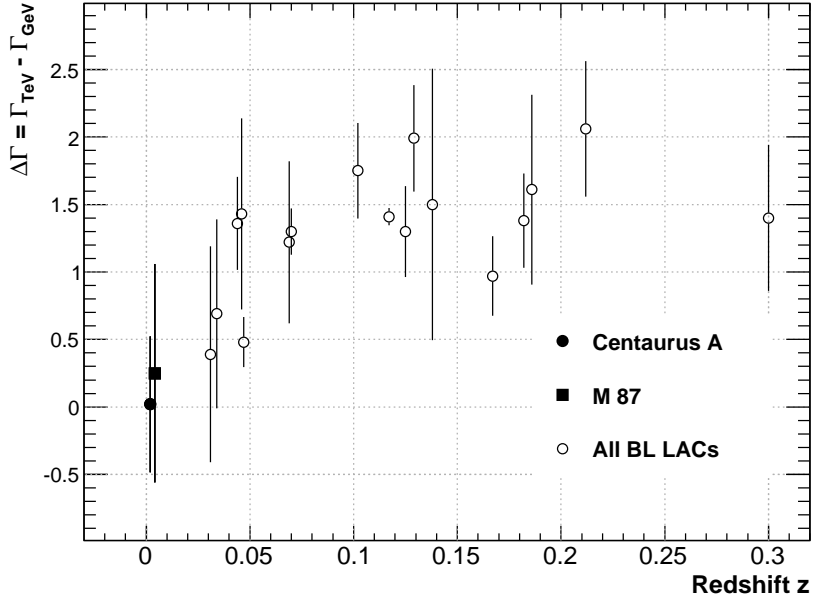,width=5in}
\end{center}
\caption{The evolution of the difference between HE and VHE photon indexes
  (Ref.~\refcite{fermi1lac}), showing that the spectral change becomes
  increasingly more important the more distant the observed source is.}
\label{fig5}
\end{figure}

It seems however that, at least for distant BL Lacs ($z\geq0.2$), most of the
VHE spectra contain photons up to hundreds of GeV, but rarely up to TeV
energies. Interestingly, extending the \textit{Fermi} spectrum by (just) one
decade, along with a conservative EBL attenuation model, reproduces the VHE
spectra quite well. This could mean that, for most of those hard \textit{Fermi}
sources, the \textit{observed} SED peak is not necessarily the
\textit{intrinsic} luminosity peak, which could be in some cases be located at
energies well above $1\,{\rm TeV}$. This can be seen in Figure~\ref{fig5}, where
a clear evolution of the observed spectral break with the redshift is possibly a
model-independent signature of the increasing effect of EBL attenuation, further
complicating the characterization of the intrinsic emission. 

\subsection{Time-of-flight measurements}

If it was well known that high-energy $\gamma$-rays can be an interesting probe
of the contents of the space they travel through, that they can also be a probe
of space \textit{itself} is a rather new and at least as interesting
development. It appears that some models for quantum gravity predict an
energy-dependence on the speed of light through Lorentz invariance
violation\cite{camelino}, which can be probed when high-energy gamma-rays travel
over cosmological distances. This has in turn been searched for by quantifying
possible energy-dependent lags in AGN light curves (assuming no intrinsic
time-effect cause), which has provided some of the most constraining upper limits to the
energy scale at which this happens (for a review see, e.g.,
Ref.~\refcite{rwag}).

\section{Prospectives}

Besides high expectations for the coming years of \textit{Fermi} results, The
ACT systems have also upgrading plans which will improve the VHE knowledge of
the sky. The MAGIC collaboration is now operating a second large telescope, and the
HESS collaboration is building a central large telescope which will operate with
the 4 smaller ones. A major change will come from the construction of larger
Cherenkov observatories such as the Cherenkov Telescope Array (CTA) or the
Advanced gamma-ray Imaging System (AGIS), with sensitivities improving by an
order of magnitude at $\sim 1\,{\rm TeV}$ and a lower energy threshold. However,
in the most optimistic case, this is not likely to happen before the next Marcel
Grossmann meeting, which will be a more appropriate venue to discuss these
developments in further details.

\end{document}